\documentclass[aps,prd,onecolumn,superscriptaddress,nofootinbib,showpacs]{revtex4}
\usepackage{epsfig,amsfonts,amsthm}
\usepackage{amsmath,amssymb}
\usepackage[normalem]{ulem}
\usepackage{color}
\usepackage[pdfborder={0 0 0}]{hyperref}
\usepackage[utf8]{inputenc}
\newcommand{\be}{\begin{equation}}
\newcommand{\ee}{\end{equation}}
\newcommand{\bea}{\begin{eqnarray}}
\newcommand{\eea}{\end{eqnarray}}

\newcommand{\doublet}[2]{ \left( \begin{array}{c}#1 \\ #2 \end{array}\right) }

\newcommand{\Z}{\mathbb{Z}}
\newcommand{\mmatrix}[4]{ \left(\! \begin{array}{ccc}#1 & #2 \\ #3 & #4 \end{array}\!\right) }
\newcommand{\mmmatrix}[9]{ \left(\! \begin{array}{ccc}#1 & #2 & #3\\ #4 & #5 & #6\\ #7 & #8 & #9\\ \end{array}\!\right) }

\newcommand{\toCP}{\xrightarrow{CP}}
\newcommand{\unit}{1\!\!1}

\newcommand{\lrpartial}{\,\partial^{\hspace{-7pt}\raise3pt\hbox{\small $\leftrightarrow$}}\!}

\def\lsim{\mathrel{\rlap{\lower4pt\hbox{\hskip1pt$\sim$}}
    \raise1pt\hbox{`$<$}}}         
\def\gsim{\mathrel{\rlap{\lower4pt\hbox{\hskip1pt$\sim$}}
    \raise1pt\hbox{$>$}}}         

\begin{document}
\title{
{\normalsize \hfill CFTP/18-007} \\*[7mm]
Multi-Higgs models with $CP$-symmetries of increasingly high order}

\author{Igor P. Ivanov}\thanks{E-mail: igor.ivanov@tecnico.ulisboa.pt}
\affiliation{CFTP, Departamento de F\'{\i}sica,
Instituto Superior T\'{e}cnico, Universidade de Lisboa,
Avenida Rovisco Pais 1, 1049 Lisboa, Portugal}
\author{Maxim Laletin}\thanks{E-mail: maxim.laletin@uliege.be}
\affiliation{Space sciences, Technologies and Astrophysics Research (STAR) Institute, Universit\'{e} de Li\`{e}ge, B\^{a}t B5A, Sart Tilman, 4000 Li\`{e}ge, Belgium}

\begin{abstract}
When building $CP$-symmetric models beyond the Standard Model,
one can impose $CP$-symmetry of higher order. 
This means that one needs to apply the $CP$-transformation more than two times to get the identity transformation,
but still the model is perfectly $CP$-conserving.
A multi-Higgs-doublet model based on $CP$-symmetry of order 4,
dubbed CP4, was recently proposed and its phenomenology is being explored.
Here, we show that the construction does not stop at CP4. We build examples
of renormalizable multi-Higgs-doublet potentials which are symmetric under CP8 or CP16, 
without leading to any accidental symmetry.
If the vacuum conserves $CP$-symmetry of order $2k$, then
the neutral scalars become $CP$-eigenstates, which are characterized not by $CP$-parities
but by $CP$-charges defined modulo $2k$.
One or more lightest states can be the DM matter candidates, which are protected against decay
not by the internal symmetry but by the exotic $CP$.
We briefly discuss their mass spectra and interaction patterns for CP8 and CP16.
\end{abstract}


\maketitle

\section{Introduction}

The Standard Model (SM) is agnostic about the origin of the $CP$-violation which we observe in weak interactions
\cite{book}.
SM simply postulates it and describes it via the complex Cabibbo-Kobayashi-Maskawa (CKM) matrix \cite{Kobayashi:1973fv}, 
but it provides no answer why the $CP$-symmetry should be broken at all.
The search for a dynamical reason of why $CP$ is broken is one of the motivations for building models
beyond the SM (bSM), especially those with non-minimal Higgs sectors \cite{Ivanov:2017dad}.
In fact, in the same year as Kobayashi and Maskawa put forward the idea that three quark generations
can accommodate all $CP$-violating phenomena \cite{Kobayashi:1973fv},
T.~D.~Lee proposed in \cite{Lee:1973iz} the two-Higgs-doublet model (2HDM), where the $CP$-symmetry is broken
spontaneously, as a result of the minimization of a $CP$-symmetric Higgs sector.
At present we know that the CKM paradigm is indeed at work, while the Higgs boson properties revealed by the LHC
are compatible with the SM Higgs \cite{Khachatryan:2016vau}.
Still, since the origin of the complex CKM matrix remains unexplained and since baryogenesis calls for yet additional
sources of $CP$-violation, the intensive exploration of 2HDM \cite{Branco:2011iw} and more sophisticated 
multi-Higgs models \cite{Ivanov:2017dad,Ilnicka:2018def} continues at full speed.

This research brought up an important technical challenge:
it is not immediately clear if a given Higgs sector is $CP$-conserving or $CP$-violating,
and whether $CP$-violation is introduced explicitly in the potential or happens spontaneously,
after its minimization.
Any phenomenological insight about $CP$-violation in the real world runs into these technical questions.
They must be clarified for a safe application to phenomenology and, therefore, 
represent an important quest on their own.

One aspect is that a model may be written in a basis in the space of Higgs doublets
which hides the presence of a $CP$-symmetry, which calls for basis-independent criteria
for $CP$-conservation or violation.
Another, somewhat surprising finding is that there may exist different forms of $CP$-{\em conservation}.
They are represented by $CP$-symmetries which cannot be related to each other via
any basis change and which lead to manifestly distinct multi-Higgs models.
They highlight the well known but, arguably, not fully appreciated fact that
irremovable complex coefficients in the lagrangian do not always indicate the presence of $CP$-violation.

The first example of such multi-Higgs-doublet models with exotic form of $CP$-conservation
was constructed in  \cite{Ivanov:2015mwl}.
Dubbed CP4 3HDM, this model is based on three Higgs doublets 
and incorporates a generalized $CP$-symmetry of order 4 denoted CP4.\footnote{We remind the reader that the order of a transformation
shows how many times one needs to apply this transformation to obtain the identity transformation.}
Although it is known since long ago that models with several scalar fields $\phi_i$, $i = 1,\dots,N$ with identical quantum numbers
allow for unconventional definitions of $CP$-symmetry \cite{Ecker:1987qp,Grimus:1995zi,Weinberg:1995mt},
\be
\phi_i(\vec x, t) \toCP {\cal CP}\,\phi_i(\vec x, t)\, {\cal CP}^{-1} = X_{ij}\phi_j^*(-\vec x, t), \quad X_{ij} \in U(N)\,.
\label{GCP}
\ee
in the vast majority of cases these definitions can be reduced to the standard one, with $X_{ij} = \delta_{ij}$, by a basis change.
For example, in 2HDM, one can define the $CP$-symmetry in the scalar sector in a variety of ways 
\cite{Ivanov:2005hg,Nishi:2006tg,Maniatis:2007de,Ferreira:2009wh}, but whatever definition one takes,
the scalar sector of the model contains, in an appropriate basis, 
the conventional $CP$-symmetry \cite{Gunion:2005ja}.\footnote{It is worth mentioning that the scalar sector 
of 2HDM can accommodate a $CP$-transformation which {\em cannot} be transformed into the usual $CP$ by any basis change
\cite{Ivanov:2005hg,Nishi:2006tg,Maniatis:2007de,Ferreira:2009wh}.
This $CP$-transformation is defined by exactly the same matrix $X$ as the one used in CP4 3HDM.
However, within 2HDM, its effective order is not 4 but 2 due to the $U(1)_Y$ rephasing symmetry.
This is best seen in the geometric picture where this unusual $CP$-transformation is described by a point reflection
rather than plane reflection in the bilinear space, see detailed discussion in \cite{Maniatis:2007de}.
It is still a {\em reflection}, that is, a transformation of order 2, but it is different from the usual $CP$.
The 2HDM based on this symmetry, which was dubbed in \cite{Maniatis:2007de} the maximally $CP$-symmetric model,
has a very peculiar phenomenology, especially when this symmetry is extended to the fermionic sector, \cite{Maniatis:2009vp}.
It turns out, however, that the imposition of this maximal $CP$-symmetry entails other symmetries in 2HDM including the usual $CP$.
The CP4 3HDM example is free from these accidental symmetries.}

However, CP4 being a symmetry transformation of order 4, is markedly different.
One needs to apply it four times, not twice, to obtain an identity transformation on fields.
Thus, it cannot be reduced to the ordinary $CP$-symmetry by any basis change.
In other words, the matrix $X_{ij}$ in \eqref{GCP} cannot be linked to $\delta_{ij}$
by any basis change. This feature has clearly visible consequences in the scalar potential:
despite the model is $CP$-conserving, it is impossible to find a basis in which all coefficients
would be real. Technically, this is due to the existence of gauge-invariant non-hermitian 
combinations of Higgs fields which are invariant under CP4 instead of being mapped to their hermitian conjugates.

This property may also be linked to an interesting group-theoretical observation 
made in \cite{Chen:2014tpa,Fallbacher:2015rea}.
If one starts with a certain symmetry group $G$ (which may include the Lorentz and gauge groups) 
and enlarges it with a $CP$-type symmetry,
then this $CP$-transformation acts on $G$ by an outer automorphism \cite{Grimus:1995zi}.
It turns out that the structure of the group $G$ and the properties of its Clebsch-Gordan coefficients
may influence this construction. In particular, for certain groups, this construction is possible but leads to 
a higher-order $CP$-transformation due to the complex Clebsch-Gordan coefficients.
This offers another look at how irremovable complex coefficients may arise in $CP$-conserving models.

CP4 3HDM is the minimal multi-Higgs-doublet model whose scalar sector incorporates only CP4 without
any accidental symmetries \cite{Ivanov:2015mwl,Ivanov:2011ae}.
If CP4 is conserved at the minimum of the Higgs potential, then the model produces two mass-degenerate scalar
dark matter (DM) candidates $h$ and $a$.
The model then resembles an enhanced version of the Inert doublet model (IDM)
\cite{Ma:2006km,Barbieri:2006dq,LopezHonorez:2006gr,Ilnicka:2015jba,Belyaev:2016lok},
with two inert doublets and with the DM candidates stabilized not by the $\Z_2$-symmetry but by a $CP$-symmetry, 
albeit an unusual one.\footnote{An example of models in which $P$-symmetry stabilizes a fermionic DM
was presented in \cite{Kuchimanchi:2012te} and appeals to the known fact that a Majorana
fermion picks up an $i$-factor upon $P$-transformation. In our case, CP4 stabilizes scalar DM candidates, 
and this phenomenon has a different origin.}
We stress that the pairwise mass-degenerate spectrum of the inert scalars arises in this model as a result not
of a continuous symmetry group, as for example in \cite{Huang:2015wts},
but of a discrete symmetry group degenerated by a peculiar $CP$-transformation.
This resemblance between CP4 3HDM and IDM is not limited to the mass spectrum but 
extends to the entire lagrangian and allows one, for example, to relate the DM dynamics of the two models.

Although the model is truly $CP$-conserving, one cannot classify $h$ and $a$ as being $CP$-even or $CP$-odd,
as they transform under CP4 as $h \toCP -a$ and $a\toCP h$.
However, one can combine them into a single complex field $\varphi$,
which then transforms under CP4 as
\be
\varphi(\vec x, t) \toCP i\varphi(-\vec x, t)\,.\label{J-eigenstates}
\ee
The presence of the $i$ factor and the absence of complex conjugation usually associated with a $CP$-transformation
are highly peculiar and were discussed at length in \cite{Aranda:2016qmp}.

With conserved CP4, one can quantify $CP$-properties of a field not by its $CP$-parity
but by a global quantum number $q$ defined modulo 4.
One then assigns $q=+1$ to $\varphi$ and $q=-1$ to its conjugate.
In any transition between initial and final states with definite $q$, 
this quantum number is additively conserved modulo 4.
When rewriting the inert self-interaction potential in terms of fields $\varphi$, terms such as $\varphi^4 + (\varphi^*)^4$ are allowed, since they also conserve the $CP$-charge $q$.

Can one go beyond CP4, while still keeping the interactions renormalizable?
Can one build a multi-Higgs model invariant under a $CP$-symmetry of order $2k$, where $k > 2$?
On the one hand, one can certainly define generalized $CP$-symmetries of an arbitrary even order.
However, if $k$ contains any prime factor other than 2, one can split the group $\Z_{2k}$ into
a pure family symmetry group and a group generated by a {\em smaller-order} $CP$-transformation.
For example, since $\Z_6 \simeq \Z_2 \times \Z_3$, imposing a $CP$-symmetry of order 6
would produce a model with a usual $CP$ and a $\Z_3$ family-symmetry group.
The only way to prevent it is to take the order of the $CP$-symmetry $2k = 2^p$, with integer $p\geq 1$. 
The usual $CP$, which is of order two, can be denoted as CP2, the first non-trivial higher-order $CP$-symmetry is CP4, 
the next ones are CP8, CP16, and so on.

Next, although one can define CP8 or CP16 transformations in multi-Higgs models and impose them on the potential,
it may easily happen that the model leads to accidental symmetries.
 For example, this is what happens in 3HDM \cite{Ivanov:2011ae}. 
Trying to impose CP8 leads to a model with an accidental continuous symmetry $U(1)$
and the usual $CP$, so that CP8 plays no special role in it.
Since the classification performed in \cite{Ivanov:2011ae} was exhaustive, it means that one needs to move beyond 
three Higgs doublets, at least as long as one keeps the renormalizability.
Thus, even though there seems to be no obstacles {\em a priori}, one should demonstrate explicitly
how such models based on yet higher-order $CP$-symmetries can be built, and what novel features they involve.

In this paper we perform this task. We begin by explaining why 3HDM and 4HDM
offer insufficient freedom to incorporate $CP$-symmetries beyond CP4,
and then build two examples of five-Higgs-doublet models based on $CP$-symmetries CP8 and CP16.
Assuming that these symmetries are respected by the minimum, we derive scalar mass spectrum
and discuss the properties of the DM candidates.
The five doublets are grouped in a natural way: one Higgs doublet acquires the vacuum expectation value (vev)
and produces the SM-like Higgs particle, while the inert sector includes two pairs of two doublets,
with the $CP$-transformation mixing the doublets within each pair.
When constructing these examples, we will explain the strategy of building models
with even higher-order $CP$-symmetries, should an interest in such models appear.

We admit that, at present, these models seem rather exotic, especially given that CP4 3HDM
phenomenology has not yet been explored in sufficient detail.
Our main motivation is purely theoretical: we want to demonstrate to the model-building community
that there exist other self-consistent options for defining $CP$-symmetry and, therefore, other routes to $CP$-violation.
It may happen that such a symmetry can arise as the residual low-energy symmetry of yet another
highly symmetric construction. Or an exotic $CP$-symmetry may single-handedly 
lead to a phenomenologically attractive and predictive model of fermion properties. 
These issues remain to be investigated in detail. For the moment, we want to convey to the community
the message that all these previously overlooked possibilities exist.

\section{NHDMs with higher order CP}

\subsection{The freedom of defining $CP$-symmetries}
\label{section:CP-freedom}

A self-consistent local quantum field theory does not uniquely specify how discrete symmetries,
such as $C$ and $P$, act on field operators \cite{Feinberg-Weinberg,Lee:1966ik,book,Weinberg:1995mt}.
There is freedom in defining these transformations, which becomes especially large
in the case of several fields with equal quantum numbers. These fields are
not physical by themselves; any linear combination of those fields which preserves
the kinetic terms will be equally acceptable as a basis choice for the theory.
Therefore, any symmetry of the Lagrangian which is supposed to incorporate a physically measurable
property, is defined up to an unconstrained basis change.

Focusing on several scalar fields $\phi_i$, $i = 1, \dots, N$ with equal quantum numbers,
one can define the $CP$-transformation as in \eqref{GCP}.
If there exists a unitary matrix $X$ such that the Lagrangian and the vacuum of a model are invariant under this 
transformation, then the model is $CP$-conserving in the very traditional sense that all $CP$-odd observables are zero, 
and the transformation \eqref{GCP} plays the role of ``the $CP$-symmetry'' of the model \cite{book}. 
It is only when {\em none} of transformations \eqref{GCP} is a symmetry of the model
that  we say that $CP$-violation takes place.

Using the basis change freedom, it is possible
to bring the matrix $X$ to a block-diagonal form \cite{Ecker:1987qp,Weinberg:1995mt}, 
which has on its diagonal either unit entries or $2\times 2$ matrices of the following type:
\be
\mmatrix{c_\alpha}{s_\alpha}{-s_\alpha}{c_\alpha}\quad \mbox{as in Ref.~\cite{Ecker:1987qp},}\quad \mbox{or}\quad
\mmatrix{0}{e^{i\alpha}}{e^{-i\alpha}}{0}\quad \mbox{as in Ref.~\cite{Weinberg:1995mt}.}\label{block}
\ee
This is the simplest form of $X$ one can achieve with basis transformations in the scalar space $\mathbb{C}^N$.

Applying the transformation \eqref{GCP} twice, one obtains a pure family transformation $a = XX^*$.
If $X$ contains at least one $2\times 2$ block with $\alpha \not = 0$ or $\pi$, then $a \not = \delta_{ij}$,
which means that the $CP$-transformation \eqref{GCP} is not an order-2 transformation. If $k$ is the smallest
integer such that $a^k = \delta_{ij}$, then we get the $CP$-transformation of order $2k$, which we denote CP$2k$, 
and the resulting family symmetry group $\Z_k$, which is generated by $a$, the square of the $CP$-transformation.
As we explained in the introduction, in order to avoid accidental symmetries, one needs to consider only $2k = 2^p$.

\subsection{The strategy}

Before moving to specific examples, let us first outline the strategy of building $N$-Higgs-doublet models 
whose only symmetries in the scalar sector are $CP$-symmetries of orders $2k$ and their powers. 

One starts by writing the Higgs potential as a sum of rephasing-invariant and rephasing-sensitive parts, $V = V_0 + V_1$.
The rephasing-invariant part can be generically written as
\be
V_0 = \sum_i m_{ii}^2 \phi_i^\dagger \phi_i + \sum_{i\le j}  \lambda_{ij} (\phi_i^\dagger \phi_i) (\phi_j^\dagger \phi_j) +
\sum_{i < j} \lambda'_{ij} (\phi_i^\dagger \phi_j) (\phi_j^\dagger \phi_i)\,,\label{V0}
\ee
with all the coefficients being real. The rephasing-sensitive part $V_1$ contains only those quadratic and quartic combinations
which are invariant under the rephasing transformation $a$.

Although for small values of $k$ and $N$, the phase-sensitive part of the potential
can be quickly constructed by trial-and-error, there exists 
an algorithmic procedure described in \cite{Ivanov:2011ae} which allows one to build $V_1$ 
for a chosen rephasing symmetry group $\Z_k$ with a given number of doublets $N$.
Of course, not all discrete groups can be implemented.
In the same work \cite{Ivanov:2011ae}, it is proven that, staying with $N$ doublets and renormalizable potentials,
one can implement cyclic groups $\Z_k$ of order $k \le 2^{N-1}$.
Trying to impose any symmetry whose order is larger than this bound unavoidably leads to accidental continuous symmetries. 
Thus, the order of generalized $CP$-symmetry in NHDM cannot exceed $2k = 2^N$.

Even if the potential $V_1$ is constructed with guess, one can always find its full rephasing symmetry group
via the systematic procedure based on Smith normal forms, which was developed in \cite{Ivanov:2011ae} 
and explained in less technical fashion in \cite{Ivanov:2013nla}.
This computation can be done by hand or implemented in a computer-algebra code.
It is in this way that one verifies the absence of accidental rephasing symmetries.
The absence of other symmetries beyond rephasing ones is guaranteed by the fact
that all free parameters in \eqref{V0} are independent.

Next, having the potential invariant under $\Z_k$ generated by $a$, one needs to check
what additional conditions on its parameters one must impose to make it invariant under
the desired $CP$-symmetry of order $2k$. 
Since higher-order $CP$-transformations mix pairs of doublets, 
there arise obvious conditions on the parameters of $V_0$ such as $m_{22}^2 = m_{33}^2$, etc.
In addition, the parameters of $V_1$ are also constrained. These constraints can be analyzed term by term.

However, instead of such analysis, we will proceed in a more efficient way.
We will first construct all bilinear combinations $\phi_i^\dagger \phi_j$ and classify them according to
their $CP$-charge $q$ defined modulo $2k$. Within each sector with definite $q$, there may exist
several bilinears $r_a$, all of them transforming in the same way under $CP$:
\be
r_a \toCP \eta^q\cdot r_a\,.\label{ra}
\ee 
It is sufficient to list only bilinears with $0 \le q \le k$; 
the complex conjugated bilinears $r_a^\dagger$ with $CP$-charges $-q$ will fill all other charge assignments from $k$ to $2k$.
In terms of these bilinears, the total potential can be schematically written as
\be
V = M_a r_a + \Lambda_{ab} r_a r_b^\dagger \,,\label{Vrarb}
\ee
where the non-zero coefficients $M_a$ span only those $r_a$ with $q=0$,
and the hermitian matrix $\Lambda_{ab}$ is block-diagonal,
with unconstrained blocks within each $q$ sector.

Once again, it is important to check that the resulting CP$2k$-invariant potential
does not acquire any accidental symmetries.
The rephasing symmetry group can again be unambiguously found 
with the Smith normal form technique \cite{Ivanov:2011ae,Ivanov:2013nla}.
The absence of the usual $CP$-symmetry is guaranteed by the fact that the hermitian matrix
$\Lambda_{ab}$ in \eqref{Vrarb} cannot be made real by any basis change.
Absence of other accidental symmetries beyond rephasing is assured by the fact that 
the matrix $\Lambda_{ab}$ has sufficiently many independent free parameters.

Since in this work we do not aim at producing {\em minimal} models but rather look for {\em examples}
of $CP$-protected scalar dark matter candidates, we will make sure that it is possible to conserve this symmetry
upon minimization of the Higgs potential. This will lead us, both for CP8 and CP16, to models with five Higgs doublets:
one SM-like $\phi_1$ and four inert ones $\phi_i$, $i = 2,3,4,5$. These inert doublets form two pairs, $(\phi_2, \phi_3)$
and $(\phi_4, \phi_5)$, which get mixed by the CP8 or CP16-transformation.
In each case, we will take the $CP$-conserving vev alignment $v_1 = v$, $v_{2,3,4,5} = 0$, 
expand the potential around the minimum, and calculate the neutral and charged scalar mass matrices.
We will confirm the general observation that the physical scalar fields in the inert sector are pairwise mass-degenerate,
just as in CP4 3HDM. For neutral scalars, we will combine pairs of real mass eigenstates into complex neutral fields
with definite $CP$-charge $q$ and briefly discuss the emerging self-interaction pattern.

\subsection{3HDM is not enough}\label{subsection-3HDM-not-enough}

It is instructive to begin the study by demonstrating why 3HDM fails to accommodate the CP8-symmetry \cite{Ivanov:2011ae}.
According to the general strategy, one first needs to write a model with rephasing symmetry $\Z_4$ 
and then extend the symmetry to CP8. 
For three Higgs doublets $\phi_i$, $i=1,2,3$, the $\Z_4$ group of symmetries is generated by 
the transformation $a_4$ which, after an appropriate basis change, can be represented as
\be
a_4 = \mmmatrix{1}{\cdot}{\cdot}{\cdot}{-i}{\cdot}{\cdot}{\cdot}{i}\,.\label{a4-3HDM}
\ee
Here, dots stand for the zero entries. The Higgs potential is written as $V= V_0 + V_1$, 
where $V_0$ given in \eqref{V0} and while the phase-sensitive part $V_1$
\be
V_1 = \lambda (\phi_2^\dagger \phi_3)^2 + \lambda' (\phi_1^\dagger \phi_2)(\phi_1^\dagger \phi_3) + h.c.
\label{VZ4-3HDM}
\ee
Here, both coefficients can be complex and must be non-zero. If at least one of them is zero,
then the number of independent phase-sensitive terms drops below $N-1$, 
and the potential acquires a continuous rephasing symmetry \cite{Ivanov:2011ae}.

We now want to require that this potential be invariant under CP8, which is generated by
$\phi_i \toCP X_{ij} \phi_j^*$ of order 8. The matrix $X$ can be brought by a basis change to the form
\be
X = \mmmatrix{1}{\cdot}{\cdot}{\cdot}{\cdot}{\eta^*}{\cdot}{\eta}{\cdot}\,,
\quad \eta \equiv e^{i \pi/4}\,, \quad \eta^8 = 1\,.\label{CP8-3HDM}
\ee
One immediately checks that applying CP8 twice produces $XX^* = a_4$ from Eq.~\eqref{a4-3HDM}.
Since CP8 mixes the doublets $\phi_2$ and $\phi_3$, one must equate their respective coefficients in $V_0$.
In addition, one requires that $V_1$ stays invariant under CP8. 
Straightforward algebra shows that under CP8
\be
(\phi_2^\dagger \phi_3)^2 \toCP \eta^4 (\phi_2^\dagger \phi_3)^2 = - (\phi_2^\dagger \phi_3)^2 \,.
\ee
Therefore, one must set $\lambda = 0$ to assure CP8-invariance of $V_1$.
Since we are left with only one rephasing-sensitive term, the potential acquires
a continuous $U(1)$ rephasing symmetry. 
Therefore, the true symmetry content of the resulting model is not the discrete group generated by CP8
but the continuous group of arbitrary phase rotations and the {\em usual} CP-transformation.
Colloquially speaking, 3HDM potential does not offer enough room to incorporate CP8
without producing accidental symmetries.

In the Appendix we show that this observation generalizes to NHDM with any $N$.
If one takes the largest cyclic group possible for NHDM, $\Z_k$ with $k = 2^{N-1}$, 
and calculates for all Higgs doublets $\phi_i$ their $q_i$ charges associated with the rephasing group $\Z_k$,
then one finds a very characteristic pattern of these charges, which involves successive powers of $2$.
However, if one starts with a $CP$-symmetry of order $2k$, then one arrives at the same symmetry group
$\Z_k$ with a very distinct pattern of charges: for any doublet with charge $q_i$
there exists a doublet with charge $-q_i$. 
These two patterns do not match.
It means that trying to impose $CP$-symmetry of order $2k=2^N$ on NHDM leads to a continuous symmetry,
which ruins the construction.

\subsection{Nor is 4HDM}\label{subsection-3HDM-not-enough}

Since CP8 requires going beyond three Higgs doublet, 
it is logical to try implementing it in 4HDM.
When building such a model, one has some freedom in constructing the corresponding matrix $X$.
In particular, one can assume that, after an appropriate basis change, 
it takes one of the following forms:
\be
X = 
\left(\begin{array}{cccc}
1 & \cdot & \cdot & \cdot \\
\cdot & 1 & \cdot & \cdot \\
\cdot & \cdot & \cdot & \eta^* \\
\cdot & \cdot & \eta & \cdot 
\end{array}
\right)
\quad \mbox{or}\quad
\left(\begin{array}{cccc}
\cdot & -i & \cdot & \cdot \\
i & \cdot & \cdot & \cdot \\
\cdot & \cdot & \cdot & \eta^* \\
\cdot & \cdot & \eta & \cdot 
\end{array}
\right)
\quad \mbox{or}\quad
\left(\begin{array}{cccc}
\cdot & \eta^* & \cdot & \cdot \\
 \eta & \cdot & \cdot & \cdot \\
\cdot & \cdot & \cdot & \eta^* \\
\cdot & \cdot & \eta & \cdot 
\end{array}
\right)\,, \quad \quad \eta^8 = 1\,.
\label{CP8-4HDM}
\ee
We have attempted constructing the corresponding models
and have found that the either assignment leads to a continuous accidental symmetry.

In the case of a single $2\times 2$ block, as in the first two matrices in Eq.~\eqref{CP8-4HDM},
we run into the same obstacle as outlined above for 3HDM. 
Thus, one must use at least two $2 \times 2$ blocks and arrange for their cross-couplings.
However, in that case, when taking the fourth power of the $CP$-transformation,
one arrives at diag$(-1,\, -1,\, -1,\, -1)$, which is identity up to the overall hypercharge transformation.
Thus, in what concerns physical consequences, the $CP$-symmetry imposed has order 4 not 8.
From the arguments similar to those described in the Appendix,
we conclude that, once again, there is a mismatch between the imposed form of the $CP$-symmetry and
the requirement that, when squared, it should generate a rephasing symmetry of order 4.
Thus, to properly impose CP8 in NHDM without generating accidental symmetries, one must go to five Higgs doublets.

\section{Building 5HDMs with CP8}

\subsection{5HDM with CP8}

The five-Higgs-doublet model 5HDM can incorporate cyclic groups with order up to 16.
By the arguments exposed in the Appendix, the maximal cyclic symmetry $\Z_{16}$ cannot be extended to CP32.
However, 5HDMs with CP8 and CP16 are well possible, and  
in this and the next sections, we construct such models.

The 5HDM uses $N=5$ Higgs doublets $\phi_i$, all with the same gauge quantum numbers.
Similarly to the previously considered 3HDM case, we define, in the appropriate basis,
the generator $a_4$ of the group $\Z_4$ and the matrix $X$ which defines CP8:
\be
a_4 = \left(\! \begin{array}{ccccc}
1 & \cdot & \cdot & \cdot & \cdot \\
\cdot & -i & \cdot & \cdot & \cdot \\
\cdot & \cdot & i & \cdot & \cdot \\
\cdot & \cdot & \cdot & -i & \cdot \\
\cdot & \cdot & \cdot & \cdot & i \\
\end{array}\!\right)\,, \qquad
X = \left(\! \begin{array}{ccccc}
1 & \cdot & \cdot & \cdot & \cdot \\
\cdot & \cdot & \eta^* & \cdot & \cdot \\
\cdot & \eta & \cdot & \cdot & \cdot \\
\cdot & \cdot & \cdot & \cdot & \eta^* \\
\cdot & \cdot & \cdot & \eta & \cdot \\
\end{array}\!\right)\,,\label{CP8}
\ee
with the same $\eta \equiv e^{i \pi/4}$.
The relation $a_4 = XX^*$ still holds.
The scalar potential can again be written as a sum of phase-invariant and phase-sensitive parts $V = V_0 + V_1$,
with $V_0$ as in \eqref{V0}, where one implicitly assumes 
that the coefficients $m_{ii}^2$, $\lambda_{ij}$, and $\lambda_{ij}'$
respect the symmetry under the simultaneous exchange $\phi_2 \leftrightarrow \phi_3$
and $\phi_4 \leftrightarrow \phi_5$.
The phase-sensitive part is now much richer than in 3HDM due to the fact
that we have two $2\times 2$ blocks in the definition of $X$.

Following the strategy outlined in the previous section, we write down all $N^2=25$ gauge-invariant bilinears $\phi_i^\dagger \phi_j$
and build out of them combinations $r_a$ which are CP8-eigenstates, that is, 
which transform under CP8 as in \eqref{ra}.
Using the shorthand notation $i \equiv \phi_i$, we list these CP8-eigenstates according to the value of $q$:
\bea
q = 0: && 1^\dagger 1, \quad 2^\dagger 2 + 3^\dagger 3, \quad 4^\dagger 4 + 5^\dagger 5,\quad
2^\dagger 4 + 5^\dagger 3, \quad 4^\dagger 2 + 3^\dagger 5, \nonumber\\
q = 1: && 2^\dagger 1 + 1^\dagger 3, \quad 4^\dagger 1 + 1^\dagger 5,\nonumber\\
q = 2: && 2^\dagger 3,\quad 4^\dagger 5, \quad 4^\dagger 3 + 2^\dagger 5, \quad 3^\dagger 4 - 5^\dagger 2 \nonumber\\
q = 3: && 1^\dagger 2 - 3^\dagger 1, \quad 1^\dagger 4 - 5^\dagger 1 \nonumber\\
q = 4: && 2^\dagger 2 - 3^\dagger 3, \quad 4^\dagger 4 - 5^\dagger 5,\quad
2^\dagger 4 - 5^\dagger 3, \quad 4^\dagger 2 - 3^\dagger 5.\label{CP8-eigenstates}
\eea
Bilinears with $CP$-charges from 5 to 8 are obtained by complex conjugating the states listed here.
Notice that only the combinations corresponding to $q = 0$ and $q = k = 4$ fall in the traditional classification
of $CP$-even/odd states. Thus, they can be coupled with other $CP$-even or odd operators of the model
in a $CP$-conserving way. The other states cannot be classified according to $CP$-parities.

In terms of these bilinears, the total potential is schematically written as in \eqref{Vrarb}.
Using the methods outlined above, one can verify that this potential indeed does not possess any other symmetry.

\subsection{Charged Higgs masses}

When minimizing the potential, we focus on the case of CP8-conserving vacuum, which implies that only the first 
doublet acquires a vev $v_1 =v$. All inert Higgs doublets are expanded as
\be
\phi_i = \doublet{H_i^+}{{ 1\over\sqrt{2}}(h_i + i a_i)}\,, \quad i = 2, 3, 4, 5.\label{expansion}
\ee
The terms of the Higgs potential which generate the scalar masses are
\bea
V &=& m_{11}^2 (1^\dagger 1) + m_{22}^2 (2^\dagger 2 + 3^\dagger 3) + m_{44}^2 (4^\dagger 4 + 5^\dagger 5) 
+ m_{24}^2 (2^\dagger 4 + 5^\dagger 3) + (m_{24}^2)^* (4^\dagger 2 + 3^\dagger 5)\nonumber\\[1mm]
&& + \lambda_1 (1^\dagger 1)^2 + \lambda_2 (1^\dagger 1)(2^\dagger 2 + 3^\dagger 3) + \lambda_3 (1^\dagger 1)(4^\dagger 4 + 5^\dagger 5) 
+ [\lambda_4 (1^\dagger 1)(2^\dagger 4 + 5^\dagger 3) + h.c.]\nonumber\\[1mm]
&&+ \lambda_5 |2^\dagger 1 + 1^\dagger 3|^2 + \lambda_6 |4^\dagger 1 + 1^\dagger 5|^2 
+ \lambda'_5 |2^\dagger 1 - 1^\dagger 3|^2 + \lambda'_6 |4^\dagger 1 - 1^\dagger 5|^2 \nonumber\\[1mm]
&&+ \left[\lambda_7 (2^\dagger 1 + 1^\dagger 3)(1^\dagger 4 + 5^\dagger 1) 
+ \lambda'_7 (2^\dagger 1 - 1^\dagger 3)(1^\dagger 4 - 5^\dagger 1) + h.c.\right].\label{V-CP8-mass-terms}
\eea
The SM-like Higgs boson acquires mass $m_h^2 = - 2m_{11}^2 = 2\lambda_1 v^2$. In the inert sector, 
we begin with the charged Higgs masses, for which only the first two lines are relevant,
and obtain the following mass terms:
\bea
&&\left(m_{22}^2 + {\lambda_2 v^2 \over 2}\right)(H_2^-H_2^+ + H_3^-H_3^+) + 
\left(m_{44}^2 + {\lambda_3 v^2 \over 2}\right)(H_4^-H_4^+ + H_5^-H_5^+) \nonumber\\[2mm]
&&\qquad + \left[\left(m_{24}^2 + {\lambda_4 v^2 \over 2}\right)(H_2^-H_4^+ + H_5^-H_3^+) + h.c.\right]\,.\label{CP8-charged-masses-1}
\eea
The charged mass matrix splits into two blocks $2\times 2$ within subspaces $(H_2^\pm, H_4^\pm)$
and $(H_3^\pm, H_5^\pm)$, with exactly the same eigenvalues in each block. Thus, the charged Higgs spectrum becomes
pairwise mass-degenerate.

Instead of explicitly diagonalizing each block, one can take one step back and simplify 
the starting potential without loss of generality. Indeed, the pairs of doublets $(\phi_2, \phi_3)$ and $(\phi_4,\phi_5)$ 
transform in exactly the same way. Therefore, one can perform basis transformations that mix $\phi_2$ and $\phi_4$
by unitary matrix $U$ and, simultaneously, $\phi_3$ and $\phi_5$ by unitary matrix $U^*$, 
and this basis change keeps the symmetry transformations $a_4$ and $X$ unchanged.
This freedom of basis change is always there, and it allows us to find such $U$ which removes the cross term
$\phi_2^\dagger \phi_4 + \phi_5^\dagger \phi_3$ altogether. 
In that basis, we still have the same potential as before, but with reparameterized coefficients. 
In particular, the charged Higgs masses will now be given only by the first line of \eqref{CP8-charged-masses-1}.
The four charged Higgses then have the following masses:
\be
m_{H_{2}^+}^2 = m_{H_{3}^+}^2 \equiv M_{H_{23}^+}^2 = m_{22}^2 + {\lambda_2 v^2 \over 2}\,,\qquad
m_{H_{4}^+}^2 = m_{H_{5}^+}^2 \equiv M_{H_{45}^+}^2 = m_{44}^2 + {\lambda_3 v^2 \over 2}\,.\label{CP8-charged-masses-2}
\ee

\subsection{Neutral Higgs masses and CP8-eigenstates}

For neutral Higgses, instead of explicitly expanding all the doublets into real components,
it is convenient to define neutral complex fields which are already CP8-eigenstates, in similarity to the states $\varphi$ and $\varPhi$ in CP4 3HDM.
These fields can be read off the table \eqref{CP8-eigenstates}; they correspond to the bilinears with $q=1$
and $q=3$ in which $\phi_1^0$ set to its vev $\langle\phi_1^0\rangle = v/\sqrt{2}$:
\bea
q=1: && \varphi_{23} = {1\over 2}(h_2 + h_3 - ia_2 + i a_3)\,,\quad \varphi_{45} = {1\over 2}(h_4 + h_5 - ia_4 + i a_5)\,,\nonumber\\
q=3: && \psi_{23} = {1\over 2}(h_2 - h_3 + ia_2 + i a_3)\,,\quad \psi_{45} = {1\over 2}(h_4 - h_5 + ia_4 + i a_5)\,,\label{CP8-phi-psi}
\eea
The two sectors corresponding to $q=1$ and $q=3$ do not mix in the mass matrix.
Staying in the charged Higgs eigenstate basis defined above, we can represent the mass terms as
\be
(\varphi_{23}^*, \varphi_{45}^*) {\cal M}_{q=1}\doublet{\varphi_{23}}{\varphi_{45}} + 
(\psi_{23}^*, \psi_{45}^*) {\cal M}_{q=3}\doublet{\psi_{23}}{\psi_{45}}\,, \label{CP8-neutral-masses-1}
\ee
where the two mass matrices are
\bea
{\cal M}_{q=1} = \mmatrix{M_{H_{23}^+}^2 + 2 \lambda_5 v^2}{2\lambda_7 v^2}{2\lambda_7^* v^2}{M_{H_{45}^+}^2 + 2\lambda_6 v^2}\,,\quad
{\cal M}_{q=3} = \mmatrix{M_{H_{23}^+}^2 + 2 \lambda'_5 v^2}{2\lambda'_7 v^2}{2\lambda_7^{\prime *} v^2}{M_{H_{45}^+}^2 + 2\lambda'_6 v^2}\,.
\label{CP8-neutral-masses-2}
\eea 
By diagonalizing them, we get the four different values for the neutral Higgs masses for the fields
$\varphi$, $\varPhi$ in the $q=1$ sector, with $m_\varphi < m_\varPhi$, and $\psi$, $\Psi$ in the $q=3$ sector, with $m_\psi < m_\Psi$.
If needed, these fields can also be written in terms of real components. In that case we have eight real fields
which are pairwise mass degenerate.

The scalars from the $q=1$ and $q=3$ sectors can interact with the $Z$-boson via $Z\varphi_i \psi_j$ vertices.
Indeed, each inert Higgs doublet $\phi_i$ produces, via its kinetic term, the term $(\bar g/2) Z_\mu (h_i \lrpartial a_i)$,
where $\bar g = \sqrt{g^2+g^{\prime 2}}$ is the combined gauge coupling and $h\lrpartial a = h (\partial_\mu a) - a (\partial_\mu h)$.
When expressed in terms of CP8-eigenstates \eqref{CP8-phi-psi}, these interaction terms become
\be
i {\bar g \over 2} Z_\mu \left( \psi_{23} \lrpartial \varphi_{23} + \psi_{45} \lrpartial \varphi_{45} 
- \psi^*_{23} \lrpartial \varphi^*_{23} - \psi^*_{45} \lrpartial \varphi^*_{45} \right)\,.\label{Z-varphii-psij}
\ee
These vertices represent the CP8-counterpart of the vertices $ZHA$ in the 2HDM and $Z\varphi\varPhi$ in the CP4 3HDM.
They conserve the CP8 quantum number: 
the sum of the internal CP8-charges of $\varphi_i$ and $\psi_i$ is $1+3=4$, which, for CP8-symmetry, is equivalent to being CP-odd.
After diagonalization of the mass matrices in the $\varphi_i$ and $\psi_i$ sectors, these interactions 
render the next-to-lightest state metastable:
\bea
&&\psi \to \varphi^* Z^{(*)} \to \varphi^* + \mbox{SM, if $m_\psi > m_\varphi$}\,,\nonumber\\
&&\varphi \to \psi^* Z^{(*)} \to \psi^* + \mbox{SM, if $m_\psi < m_\varphi$}\,.\label{psiZphi}
\eea
To avoid confusion, we stress that notation $\varphi^*$ denotes the state conjugated to $\varphi$ 
(which, contrary to the usual expectation, is {\em not} the antiparticle to $\varphi$, see detailed discussion in \cite{Aranda:2016qmp}), 
while $Z^{(*)}$ denotes a real or virtual $Z$-boson.
The only exception is when the lightest states in these two sectors are orthogonal,
which would forbid $Z\varphi\psi$-vertex and render {\em both} scalars stable.

The self-interaction in the inert sector leads to several interaction terms involving $\varphi$
and $\psi$: 
\be
V(\varphi,\psi) = 
\lambda_{\varphi} |\varphi|^4 + \lambda_{\psi} |\psi|^4 + \lambda_{\varphi\psi} |\varphi|^2|\psi|^2 +
\left[\lambda_{13}\varphi (\psi^*)^3 + \lambda_{31}\varphi^3 \psi^* + \lambda_{22}\varphi^2 \psi^2 + h.c.\right]
\ee
All coefficients here are independent; $\lambda_{13}$,  $\lambda_{31}$, and $\lambda_{22}$ can be complex,
but it does not imply $CP$-violation, because the scalars here are themselves $CP$-eigenstates, 
and the CP8-charge is conserved by each term separately.

These interactions switch on new channels for two identical DM candidates:
although the direct annihilation $\psi\psi \to$ SM is forbidden by the CP8 conservation,
the semi-annihilation processes $\psi\psi \to \psi^* \varphi$ is allowed for $m_\psi > m_\varphi$. 
Finally, for sufficiently large mass splitting, the direct triple decays are also allowed:
\be
\psi \to \varphi\varphi\varphi\, \mbox{, if $m_\psi > 3m_\varphi$}\,,\qquad
\varphi \to \psi\psi\psi\, \mbox{, if $m_\varphi > 3m_\psi$}\,.
\ee

\section{Building 5HDM with CP16}\label{section-CP16}

In this section, we build yet another version of 5HDM, the one with CP16.
We use the same strategy as before, but with the new parameter $\xi \equiv \exp(i \pi/8)$ instead of $\eta = \exp(i \pi/4)$.
The first attempt is to use the same structure for $X$ as before:
\be
a_8 = \left(\! \begin{array}{ccccc}
1 & \cdot & \cdot & \cdot & \cdot \\
\cdot & \eta^* & \cdot & \cdot & \cdot \\
\cdot & \cdot & \eta & \cdot & \cdot \\
\cdot & \cdot & \cdot & \eta^* & \cdot \\
\cdot & \cdot & \cdot & \cdot & \eta \\
\end{array}\!\right)\,, \qquad
X = \left(\! \begin{array}{ccccc}
1 & \cdot & \cdot & \cdot & \cdot \\
\cdot & \cdot & \xi^* & \cdot & \cdot \\
\cdot & \xi & \cdot & \cdot & \cdot \\
\cdot & \cdot & \cdot & \cdot & \xi^* \\
\cdot & \cdot & \cdot & \xi & \cdot \\
\end{array}\!\right)\,,\label{CP16}
\ee
and classify the bilinears according to their CP-charges $q$ defined modulo $2k=16$.
This classification is similar to Eq.~\eqref{CP8-eigenstates} but with an important change:
\bea
q = 0: && 1^\dagger 1, \quad 2^\dagger 2 + 3^\dagger 3, \quad 4^\dagger 4 + 5^\dagger 5,\quad
2^\dagger 4 + 5^\dagger 3, \quad 4^\dagger 2 + 3^\dagger 5 \nonumber\\
q = 1: && 2^\dagger 1 + 1^\dagger 3, \quad 4^\dagger 1 + 1^\dagger 5\nonumber\\
q = 2: && 2^\dagger 3,\quad 4^\dagger 5, \quad 4^\dagger 3 + 2^\dagger 5\nonumber\\
q=k-2: && 3^\dagger 4 - 5^\dagger 2 \nonumber\\
q = k-1: && 1^\dagger 2 - 3^\dagger 1, \quad 1^\dagger 4 - 5^\dagger 1 \nonumber\\
q = k: && 2^\dagger 2 - 3^\dagger 3, \quad 4^\dagger 4 - 5^\dagger 5,\quad
2^\dagger 4 - 5^\dagger 3, \quad 4^\dagger 2 - 3^\dagger 5.\label{CP2k-eigenstates}
\eea
Once again, the states with $q=0$ and $q=k$ are $CP$-even and $CP$-odd in the traditional sense.
The key difference with respect to CP8 case of Eq.~\eqref{CP8-eigenstates} is that now, with $k=8$, 
the charges $q = k-2$ and $q=2$ are distinct, and the matrix $\Lambda_{ab}$ does not mix them. 
Thus, $\Lambda_{ab}$ stays block diagonal, with blocks corresponding to subspaces of distinct values of $q$. 
But then the structure of $\Lambda_{ab}$ does not depend on the value of $k$ provided $k > 4$. 
It means that the same potential is invariant not only under CP16 but also under any higher-order CP$2k$, 
as well as under the continuous $U(1)$ transformations generated by $a_8$ in Eq.~\eqref{CP16} with
$\eta$ replaced by any phase rotation.
In short, the CP16 leads to accidental symmetries including $U(1)$ and the usual CP. 

However, we can try another quantum number assignment:
\be
a_8 = \left(\! \begin{array}{ccccc}
1 & \cdot & \cdot & \cdot & \cdot \\
\cdot & \eta^* & \cdot & \cdot & \cdot \\
\cdot & \cdot & \eta & \cdot & \cdot \\
\cdot & \cdot & \cdot & (\eta^3)^* & \cdot \\
\cdot & \cdot & \cdot & \cdot & \eta^3 \\
\end{array}\!\right)\,, \qquad
X = \left(\! \begin{array}{ccccc}
1 & \cdot & \cdot & \cdot & \cdot \\
\cdot & \cdot & \xi^* & \cdot & \cdot \\
\cdot & \xi & \cdot & \cdot & \cdot \\
\cdot & \cdot & \cdot & \cdot & (\xi^3)^* \\
\cdot & \cdot & \cdot & \xi^3 & \cdot \\
\end{array}\!\right)\,,\label{CP16-new}
\ee
Within CP8, this assignment could be reduces to the previously considered one by rephasing within the last block,
while here it leads to an essentially different model.
We also remark that it is absolutely inessential which block is the third power of which.
One can equally well denote $\eta' \equiv \eta^3$ and then observe that $\eta = \eta^9 = (\eta')^3$.
Group-theoretically, this reflects the fact that the four inert doublets are transformed under $a_8$
by the four distinct generators of the rephasing group $\Z_8$: $\eta$, $\eta^3$, $\eta^5$, and $\eta^7$.

Again, classifying the bilinear transformations for CP$2k$, we get:
\bea
q = 0: && 1^\dagger 1, \quad 2^\dagger 2 + 3^\dagger 3, \quad 4^\dagger 4 + 5^\dagger 5\nonumber\\
q = 1: && 2^\dagger 1 + 1^\dagger 3 \nonumber\\
q = 2: && 2^\dagger 3, \quad 4^\dagger 2 + 3^\dagger 5 \nonumber\\
q = 3: && 4^\dagger 1 + 1^\dagger 5 \nonumber\\
q = 4: && 4^\dagger 3 + 2^\dagger 5\nonumber\\
q = 6: && 4^\dagger 5 \nonumber\\
q=k-4: && 3^\dagger 4 - 5^\dagger 2 \nonumber\\
q=k-3: && 1^\dagger 4 - 5^\dagger 1 \nonumber\\
q=k-2: && 2^\dagger 4 - 5^\dagger 3 \nonumber\\
q = k-1: && 1^\dagger 2 - 3^\dagger 1 \nonumber\\
q = k: && 2^\dagger 2 - 3^\dagger 3, \quad 4^\dagger 4 - 5^\dagger 5.\label{CP2k-eigenstates-option2}
\eea
In the case of CP16, the value of $k=8$. Then, the CP-charges $q=k-4$ and $q=4$ are identical,
and so are the charges $q=k-2$ and $q=6$. The corresponding bilinears can be coupled via $\Lambda_{ab}$,
and the resulting potential will {\em not} have the continuous symmetry.

Using the Smith normal form technique, one also verifies that the rephasing symmetry of this model 
is indeed $\Z_8$. In order to check that the model does not accidentally acquire the usual $CP$-symmetry,
let us notice that the hermitian matrix $\Lambda_{ab}$ has three complex off-diagonal entries, 
coming from the $2\times 2$ blocks with charges $q = 2, 4, 6$.
They generate six different rephasing-sensitive terms in the potential.
Using the rephasing freedom, one can make two of these entries real, but not all three of them.
Thus, the coupling matrix $\Lambda_{ab}$ cannot be made real in any basis.
This fact forbids the usual $CP$-symmetry as well as the symmetry under 
$\phi_2 \leftrightarrow \phi_3$, $\phi_4 \leftrightarrow \phi_5$.
Thus, we have a viable 5HDM with CP16 and no accidental symmetry.

We proceed to the mass spectrum calculation for the case of unbroken CP16.
The terms in the Higgs potential that generate the scalar masses are very similar
to Eq.~\eqref{V-CP8-mass-terms} but with a few terms omitted:
\bea
V &=& m_{11}^2 (1^\dagger 1) + m_{22}^2 (2^\dagger 2 + 3^\dagger 3) + m_{44}^2 (4^\dagger 4 + 5^\dagger 5) \nonumber\\[1mm]
&& + \lambda_1 (1^\dagger 1)^2 + \lambda_2 (1^\dagger 1)(2^\dagger 2 + 3^\dagger 3) 
+ \lambda_3 (1^\dagger 1)(4^\dagger 4 + 5^\dagger 5)\nonumber\\[1mm]
&&+ \lambda_5 |2^\dagger 1 + 1^\dagger 3|^2 + \lambda_6 |4^\dagger 1 + 1^\dagger 5|^2 
+ \lambda'_5 |2^\dagger 1 - 1^\dagger 3|^2 + \lambda'_6 |4^\dagger 1 - 1^\dagger 5|^2.\label{V-CP16-mass-terms}
\eea
Therefore, we can reuse exactly the same formulas for scalar masses as before, Eqs.~\eqref{CP8-charged-masses-2} 
and \eqref{CP8-neutral-masses-2}, but just set $\lambda_7 = \lambda_7' = 0$ in the latter.
The matrices in \eqref{CP8-neutral-masses-2} become diagonal, which is to be  expected because the four complex neutral
fields carry now all distinct CP-charges:
\bea
q=1: && \varphi_{23} = {1\over 2}(h_2 + h_3 - ia_2 + i a_3)\,,\quad m^2_{\varphi_{23}} = M_{H_{23}^+}^2 + 2 \lambda_5 v^2\nonumber\\
q=3: && \varphi_{45} = {1\over 2}(h_4 + h_5 - ia_4 + i a_5)\,, \quad m^2_{\varphi_{45}} = M_{H_{45}^+}^2 + 2 \lambda_6 v^2\nonumber\\
q=5: && \psi_{45} = {1\over 2}(h_4 - h_5 + ia_4 + i a_5)\,, \quad m^2_{\psi_{45}} = M_{H_{45}^+}^2 + 2 \lambda_6' v^2\nonumber\\
q=7: && \psi_{23} = {1\over 2}(h_2 - h_3 + ia_2 + i a_3)\,, \quad m^2_{\psi_{23}} = M_{H_{23}^+}^2 + 2 \lambda_5' v^2\,.\label{CP16-phi-psi}
\eea
The interaction vertices $Z\varphi_i\psi_j$ remain as in Eq.~\eqref{Z-varphii-psij},
and they are already written in terms of mass states. All these vertices still conserve the CP16-charge $q$.
These vertices lead to decays \eqref{psiZphi} within the $23$ and $45$ subsectors, but they do not couple the two sectors,
which renders the lightest states in the two inert subsectors stable.

The self-interaction pattern in the inert sector depends on which states in the $23$ and $45$ subsectors are the lightest ones.
For example, if the DM candidates are $\varphi_{23}$ and $\varphi_{45}$, the self-interaction between them is given by
\be
V(\varphi_{23},\varphi_{45}) = \lambda_{23}|\varphi_{23}|^4 + \lambda_{45}|\varphi_{45}|^4 + 
\lambda_{2345}|\varphi_{23}|^2 |\varphi_{45}|^2 + \left[\lambda (\varphi_{23})^3 \varphi_{45}^* + h.c.\right]\,.
\ee
If it happens that the DM candidates are $\varphi_{23}$ and $\psi_{45}$, then the self-interaction terms are
\be
V(\varphi_{23},\psi_{45}) = \lambda_{23}|\varphi_{23}|^4 + \lambda_{45}|\psi_{45}|^4 + 
\lambda_{2345}|\varphi_{23}|^2 |\psi_{45}|^2 + \left[\lambda \varphi_{23} (\psi_{45})^3 + h.c.\right]\,.
\ee
In any of these cases, there exists an interaction term involving asymmetric combinations of the two fields,
which can lead to semi-annihilation and decays, in similarity to what we found in the CP8 case.

\section{Discussion and conclusions}

The key message of this study is that it is well possible to construct renormalizable multi-Higgs models 
whose symmetry content is given only by a higher-order $CP$ and its powers. 
The CP4-symmetric 3HDM proposed initially in \cite{Ivanov:2015mwl} is the simplest example of this kind,
but it is not the only possibility. We have constructed here two versions of 5HDM with CP8 and CP16,
and the methods we have used can be generalized to $CP$-symmetries of any order $2k = 2^p$, 
should the need arise.

If the vacuum respects the higher-order $CP$-symmetry, then the real scalars emerging from the inert sector
are pairwise mass-degenerate and can be grouped into complex neutral fields $\varphi_i$ which are $CP$-eigenstates.
Just as in the CP4 3HDM example, their $CP$-properties are described by CP$2k$-charges $q_i$ defined modulo $2k$.
They generalize the notion of $CP$-parity (that is, $CP$-charge defined modulo 2) for the usual $CP$-symmetry of order 2.

The lightest scalar in the inert sector serves as the DM candidate, and its stability is insured by the exotic $CP$-symmetry 
rather than internal symmetry. Models with elaborate $CP$ sectors, such as CP16 5HDM, can contain two or more 
DM candidates with different $CP$-charges.

One may ask whether there is any difference between models based on CP$2k$, considered here, 
and the more traditional multi-Higgs models based on rephasing symmetries of order $2k$,
see examples in \cite{Belanger:2012zr,Belanger:2014bga,Ivanov:2012hc}. 
Despite the symmetry group in both cases is the same, $\Z_{2k}$, there are several distinctions.

First, the CP$2k$-based models possess vertices of the type $Z_\mu \varphi_i\lrpartial \varphi_j$, where $q_i + q_j = k \not = 0$.
Such vertices lead to new coannihilation channels in the DM evolution and to novel opportunities 
to detect scalars with exotic $CP$-properties at colliders.
Such vertices are possible because the Lorentz structure of this interaction term is by itself $CP$-odd
and requires the internal $CP$-properties of the two fields $\varphi_i\varphi_j$ to organize themselves into a $CP$-odd combination.
In the usual case, for example in the $CP$-conserving 2HDM, 
the corresponding vertex is $ZHA$, where $H$ is $CP$-even and $A$ is $CP$-odd, so that their product is $CP$-odd.
In the CP$2k$-symmetric models, one just arranges $q_i + q_j = k$, which exactly corresponds to being $CP$-odd.
In the traditional $\Z_{2k}$-symmetric model, such vertices are impossible because the symmetry is {\em internal},
and therefore the $\Z_{2k}$-charges of fields in any vertex must be a multiple of $2k$.

Second, the mere fact that a complex field $\varphi$ is a $CP$-eigenstate means that it is its own antiparticle.
The one-particle state $\varphi^*|0\rangle$ is not an antiparticle to $\varphi|0\rangle$ but is rather 
a {\em different} particle with the same mass. This doubling of spectrum is only possible for zero-charge fields
and is discussed at length in \cite{Aranda:2016qmp}.

This feature allows one to consider an asymmetric DM evolution regime, in which $\varphi$ dominates over $\varphi^*$. 
However, unlike the traditional asymmetric DM models \cite{Zurek:2013wia,Petraki:2013wwa}, 
this imbalance does {\em not} imply particle-antiparticle asymmetry.
Constructing a model which exhibits such an imbalance and studying its late-time observational signatures
is a task for future investigation.

Having demonstrated that it is possible to build $CP$-conserving models based on various CP$2k$-symmetries,
one can ask whether this distinction is observable in any imaginable experiment. 
If it is, we arrive at the exciting possibility of determining experimentally the order of the $CP$-symmetry 
which the real world is closest to. This question was already posed in \cite{Ivanov:2015mwl}, but it remains unanswered.

On the theoretical side, it is interesting to see if a CP$2k$-symmetry can arise as a low-energy residual symmetry
from a more symmetric model at high energy scale, whose high symmetry spontaneously breaks down at lower energies.
All existing models of this kind generate at lower energies only usual family symmetries, not an exotic $CP$.
On the other hand, as established in \cite{Chen:2014tpa,Fallbacher:2015rea},
certain symmetry groups $G$ not only allow but even {\em require} the $CP$-symmetry to be of higher-order.
Thus, equipping a symmetric model with higher-order symmetries is not a problem; one just needs to make sure
it is the only symmetry to survive at low energies.
If such a construction leading to a residual CP$2k$-symmetry is found, it may provide a natural explanation
how the CP8 or CP16 5HDMs could emerge from a highly symmetric construction with one scalar singlet $\phi_1$
and one quadruplet $(\phi_2,\phi_3,\phi_4,\phi_5)$. It will then serve as an additional motivation
to look deeper at this exotic form of $CP$-conservation and its observable consequences.
\bigskip

We thank Roman Pasechnik and Andreas Trautner for useful discussions.
I.P.I. acknowledges funding from the the Portuguese
\textit{Fun\-da\-\c{c}\~{a}o para a Ci\^{e}ncia e a Tecnologia} (FCT) through the FCT Investigator 
contract IF/00989/2014/CP1214/CT0004 under the IF2014 Programme,
and through the contracts UID/FIS/00777/2013 and CERN/FIS-NUC/0010/2015,
which are partially funded through POCTI (FEDER), COMPETE, QREN, and the EU.
I.P.I. also acknowledges support from National Science Center, Poland, via the project Harmonia (UMO-2015/18/M/ST2/00518).
The work of M.L. is supported by a FRIA grant (F.N.R.S.).

\appendix

\section{NHDM with the maximal cyclic symmetry}

In section \eqref{subsection-3HDM-not-enough} we saw that trying to impose a $CP$-symmetry of order 8
in 3HDM leads to continuous family symmetry and a usual $CP$.
One may ask if this is a general result.
Here, we explore in some detail the $CP$ properties of the $N$-Higgs-doublet model scalar sector
with the maximal cyclic symmetry $\Z_k$, where $k = 2^{N-1}$.
We prove that it is indeed impossible to extend it to a $CP$-symmetry of order $2k = 2^N$
without producing continuous accidental symmetries.
However, the fundamental reason is slightly different from what we saw when attempting to impose CP8 in 3HDM.

We begin by reminding the reader of the result of \cite{Ivanov:2011ae} 
that the largest cyclic group which can be imposed on
the scalar potential of the $N$-Higgs-doublet model is $\Z_k$, where $k = 2^{N-1}$.
Trying to impose any larger cyclic group will unavoidably produce a model with continuous rephasing symmetry.

It is remarkable that, starting from this group-theoretical fact, one can construct the Higgs potential of this model 
in an essentially unique way, presented already in \cite{Ivanov:2011ae}.
At first glance, this may seem surprising. Indeed, one first finds a basis in which the generator of this symmetry $a_k$
acts on all doublets by rephasing. But there is a huge variety of ways $a_k$ can act on each individual doublet.
One can define such action as $\phi_i \to \exp(2\pi i q_i /k)$, and each particular implementation of $\Z_k$
is defined by its spectrum of charges $q_i$ defined modulo $k$.
Different $q_i$ spectra will produce nonequivalent models with the same $\Z_k$ symmetry 
(in fact, we already encountered this situation in section~\ref{section-CP16} when we were building CP16 5HDM).
However, we prove below that there exists, up to permutation, a unique assignment of charges, 
for which the potential does not acquire accidental continuous symmetries.

Next, we briefly recap the technique based on the Smith normal form (SNF), which was developed in 
\cite{Ivanov:2011ae} to establish the rephasing symmetry group of any potential and
the exact form of the NHDM potential with the maximal cyclic symmetry group $\Z_k$, $k = 2^{N-1}$.

For any scalar potential one first checks how each individual term changes under a generic global rephasing transformation
$\phi_j \to e^{i\alpha_j}\phi_j$. The $i$-th term picks up the phase factor $d_{ij} \alpha_j$, 
where the integer coefficients $d_{ij}$ are immediately read off the expression for the $i$-term.
For example, if the first term is $(\phi_2^\dagger \phi_1)(\phi_3^\dagger \phi_1)$,
its coefficients are $d_{1j} = (2, -1, -1, 0, \dots, 0)$.
Going through all $m$ rephasing-sensitive terms, one builds in this way the coefficient matrix $d_{ij}$, 
which is an integer-valued rectangular matrix $m \times N$.
Then one can apply a sequence of certain elementary steps and reach its Smith normal form (SNF). 
The Smith normal form exists and is unique for any rectangular matrix with integer coefficients.
The diagonal entries of the SNF immediately give the rephasing symmetry group of the potential.

The explicit form of the $\Z_k$-symmetric NHDM was given in \cite{Ivanov:2011ae}:
\be
d = \left(\begin{array}{cccccccc}
2 & -1 & 0 & 0 & \cdots & 0 & 0 & -1 \\
0 & 2 & -1 & 0 & \cdots & 0 & 0 & -1 \\
0 & 0 & 2 & -1 & \cdots & 0 & 0 & -1 \\
\vdots &  &  & & & & & \vdots \\
0 & 0 & 0 & 0 & \cdots & 2 & -1 & -1 \\
0 & 0 & 0 & 0 & \cdots & 0 & 2 & -2 
\end{array}
\right)\label{dij-max}.
\ee
Notice that each row has the following properties: $\sum_j d_{ij} = 0$,
which reflects the global overall rephasing symmetry, a subgroup of $U(1)_Y$,
and $\sum_j |d_{ij}| = 4$, which reflects the fact that all interaction terms used here are quartic.
The SNF of this matrix has on its diagonal the sequence $(1, 1, \dots, 1, 2^{N-1})$, which indicates the rephasing
symmetry group $\Z_{2^{N-1}}$, in addition to the overall $U(1)_Y$ rephasings of all doublets.
Since the SNF is unique and since its construction is invertible, 
any NHDM potential with the same symmetry group $\Z_{2^{N-1}}$ without any accidental symmetry
can be brought to this form by an appropriate basis transformation.

The Higgs potential encoded in this matrix is
\be
V_1 = \lambda_1 (\phi_N^\dagger \phi_1)(\phi_2^\dagger \phi_1) + 
\lambda_2 (\phi_N^\dagger \phi_2)(\phi_3^\dagger \phi_2) + \cdots
+ \lambda_{N-2} (\phi_N^\dagger \phi_{N-2})(\phi_{N-1}^\dagger \phi_{N-2}) 
+ \lambda_{N-1} (\phi_N^\dagger \phi_{N-1})^2 + h.c.,
\label{NHDM-max}
\ee
where all coefficients can be complex.
From this expressions one can immediately obtain the spectrum of $\Z_k$ charges:
\be
q_i \ =\ (1,\ 2,\ 4,\ \cdots,\ 2^{N-2},\ 0)\label{charges}. 
\ee
All charges are defined modulo $k = 2^{N-1}$. There is a freedom in simultaneous shift of all
charges by the same value, but their {\em differences} remain as they are.
Up to permutation, this is the only charge assignment which is compatible with $\Z_k$
and is capable of generating $N-1$ different terms, thus avoiding accidental continuous symmetries.

The potential \eqref{NHDM-max} has $N-1$ terms constructed of $N$ Higgs doublets.
There exist no other renormalizable terms invariant under the same symmetry.
This can be seen from the matrix $d$ itself.
Suppose there exists yet another term, which would appear in this matrix as $N$-th row $d_{Nj}$.
Since the matrix is of rank $N-1$, the new row can be written as a linear combination of the existing rows
with integer coefficients.
It is immediately seen by direct inspection that any such combination would produce
a row with $\sum_j |d_{Nj}| > 4$. Therefore, any such term can only be of higher order.

Suppose now we want to impose the $CP$-symmetry of order $2k$.
Denoting the generator of the cyclic group $\Z_k$ by $a$ and the generator of the CP$2k$-symmetry by $J$,
we are looking for such a construction which satisfies $J^2 = a$.
As mentioned in the introduction, a higher-order $CP$-transformation $J$ acts on doublets as $\phi_i \to X_{ij}\phi_j^*$,
where the matrix $X$ can be brought to the block-diagonal basis, 
with the diagonal containing either entries $1$ or $2\times 2$ blocks of the form \cite{Weinberg:1995mt}
\be
\mmatrix{0}{e^{i\alpha}}{e^{-i\alpha}}{0}\,.\label{block1}
\ee
When squaring the $CP$-transformation, one obtains the diagonal matrix $a = XX^*$.
Each block \eqref{block1} in $X$ contributes a pair of {\em mutually conjugate} entries $e^{\pm 2i\alpha}$ to its diagonal.
Since $a^k = \unit$, one must make sure that $\alpha$ is a multiple of $\pi/k$.
Different blocks can contain $\alpha$'s as different multiples of $\pi/k$, 
but in any case each block produces a pair of doublets with {\em opposite} charge $q_i$.
Thus, the overall $\Z_k$-charge spectrum, emerging from a CP$2k$-symmetric model,
must always exhibit a reflection symmetry: for every doublet with charge $q_i$,
there exists a doublet with charge $-q_i$.

However, we have already found the unique spectrum of $\Z_k$-charges, which does not lead to accidental symmetries,
Eq.~\eqref{charges}. That spectrum does {\em not} possess this reflection symmetry for $N> 3$.
The conclusion is that although it is possible to define a $CP$-symmetry of order $2k = 2^N$ in NHDM scalar sector,
it will produce fewer than $N-1$ rephasing-sensitive terms and, hence, the potential will contain a continuous rephasing symmetry group and a usual $CP$-symmetry. 
This proves that NHDM with maximal cyclic symmetry group $\Z_k$, with $k = 2^{N-1}$,
has so rigid structure that it cannot accommodate the discrete CP$2k$-symmetric structure.

3HDM is somewhat special. The $\Z_4$ charges are $q_i = (1, 2, 0)$ and they can be shifted by one unit to become $q_i = (0, 1, -1)$.
This spectrum indeed demonstrates the reflection symmetry mentioned. 
Therefore, one can actually construct the desired CP8-transformation whose square is the generator of $\Z_4$.
Still, the model acquires an accidental continuous symmetry as we saw in section~\eqref{subsection-3HDM-not-enough} .

We conclude this study of the NHDM scalar sector with the maximal cyclic symmetry group by noticing 
that the potential (\ref{NHDM-max}) is, in fact, {\em automatically} $CP$-invariant under a $CP$-symmetry of order 2.
Indeed, one can rephase $N$ doublets in such a way that all $N-1$ coefficients $\lambda_i$ in \eqref{NHDM-max}
become real, for an accurate proof see section~4 of \cite{Branco:2015bfb}. 
Their reality implies that the potential is invariant under the usual complex conjugation, that is $CP$-symmetry of order 2.
In addition to being explicitly $CP$-conserving, this model also forbids spontaneous $CP$-violation.
Once again, this conclusion follows from \cite{Branco:2015bfb}, where it was shown that
if a rephasing symmetry protects the model from explicit $CP$-violation, it also protects it against spontaneous $CP$-violation.
The only class of models where this relation does not hold must involve terms with four {\em different}
fields such as $(\phi^\dagger_1 \phi_2)(\phi^\dagger_3 \phi_4)$. However, such terms are absent in our case.


\end{document}